\documentclass[12pt,preprint]{aastex}

\shorttitle{Fisher Analysis on Wideband Polarimetry}
\shortauthors{Ideguchi et al.}

\begin{document}
\title{
Fisher Analysis on Wide-Band Polarimetry for Probing \\
the Intergalactic Magnetic Field
} 
\author{
Shinsuke Ideguchi$^{1}$, Keitaro Takahashi$^{1}$, Takuya Akahori$^{2}$, Kohei Kumazaki$^{3}$, and Dongsu Ryu$^{4}$
}
\affil{
$^1$Kumamoto University, 2-39-1, Kurokami, Kumamoto 860-8555, Japan; 121d9001@st.kumamoto-u.ac.jp, keitaro@sci.kumamoto-u.ac.jp\\
$^2$Sydney Institute for Astronomy, School of Physics, The University of Sydney, NSW 2006, Australia: akahori@physics.usyd.edu.au\\
$^3$Nagoya University, Furo-cho, Chikusa-ku, Nagoya 464-8601, Japan; kumazaki@a.phys.nagoya-u.ac.jp\\
$^4$Department of Astronomy and Space Science, Chungnam National
University, Daejeon, Republic of Korea; ryu@canopus.cnu.ac.kr
}

\begin{abstract}
We investigate the capability of ongoing radio telescopes for probing Faraday rotation measure (RM) due to the intergalactic magnetic field (IGMF) in the large-scale structure of the universe which is expected to be of order $O(1)~{\rm rad/m^2}$. We consider polarization observations of a compact radio source such as quasars behind a diffuse source such as the Galaxy, and calculate Stokes parameters $Q$ and $U$ assuming a simple model of the Faraday dispersion functions with Gaussian shape. Then, we perform the Fisher analysis to estimate the expected errors in the model parameters from QU-fitting of polarization intensity, accounting for sensitivities and frequency bands of Australian Square Kilometer Array Pathfinder, Low Frequency Array, and the Giant Meterwave Radio Telescope. Finally, we examine the condition on the source intensities which are required to detect the IGMF. Our analysis indicates that the QU-fitting is promising for forthcoming wideband polarimetry to explore RM due to the IGMF in filaments of galaxies.
\end{abstract}


\section{Introduction}
\label{section1}

The intergalactic magnetic field (IGMF) in clusters and filaments of galaxies has attracted much attention. A number of mechanisms which produce primordial magnetic fields in early universe have been proposed, such as inflation \citep{tw88,de09}, phase transition \citep{ho83}, density fluctuations \citep{tak05,ich06} and reionization \citep{la03,la05,ads10}. If such seed fields of any origins existed in the intergalactic medium (IGM) in early universe, they would have been amplified through compression and dynamo process caused by the hierarchical structure formation with dark matter, and affected also by activities of galaxies (e.g.,  \cite{rkb98,dt08,rkcd08,ds09,don09,xu10,xu11}). Since filaments of galaxies are the front of the IGMF evolution, explorations of the IGMF in filaments of galaxies as well as clusters of galaxies are critically important to elucidate generations of seed fields, dynamics of the IGM, and activities of galaxies.

Cosmic magnetic fields including the IGMF have been investigated through observations of Faraday rotation (for another method to probe the IGMF using gamma-rays see, e.g., \cite{tak12,tak13} and references therein.) Faraday rotation is the rotation of the polarization plane of electromagnetic waves due to magnetic fields in the propagation region. Multiple observations of the polarization angle with different wavelengths give the Faraday rotation measure (RM), which is the integration of magnetic fields along the line of sight (LOS) with a weight of electron density. Assuming the distribution of the electron density of the plasma, LOS magnetic field strength can be estimated from the RM. This method has been successfully applied to the studies of magnetic fields in external galaxies (e.g., \cite{gae05,bec09}) and in galaxy clusters (e.g., \cite{car02,gov10}).

This method, however, is not effective in the case that the intervening magnetic fields of our interest do not dominate the observed Faraday rotation. Such a situation actually arises in the case of studying RM due to the IGMF in filaments of galaxies. For instance, \cite{ar10,ar11} recently simulated that the root-mean-square (rms) value of RM due to the IGMF in filaments of galaxies is 1-several ${\rm rad/m^2}$. On the other hand, the RM due to the Galactic magnetic field, which always exist as an inevitable contamination for studying extragalactic ones, is generally much larger than those values and, even toward high galactic latitude, they are comparable \citep{mao10,sti11,arkg13}. Therefore, we need the sophisticated methods which enable us to estimate and subtract the RM from the Galaxy and/or any other contaminations.

Recent dramatic extensions of wideband polarimetry allow us to apply some new approaches. One of the promising methods to estimate the LOS distribution of RM is Faraday tomography or Faraday RM Synthesis \citep{bu66, br05}. Actually, \cite{aktr12} demonstrated that part of the RM due to the IGMF could be directly detected with future ultra-wideband polarimetry by the Square Kilometer Array (SKA) or its pathfinder/precursors such as the Australian Square Kilometer Array Pathfinder (ASKAP), LOw Frequency ARray (LOFAR) and the Giant Metrewave Radio Telescope (GMRT). Faraday tomography will be further improved with developments of decomposition techniques such as phase correction \citep{br05} and RMCLEAN (e.g., \cite{h09,bel12}).

Another promising method, which we study in this paper, is the so-called QU-fitting, where one constructs a source model and estimates the parameters of the model by fitting the data, Stokes parameters $Q$ and $U$ as functions of the wavelength, with the model. Using the method, \cite{osu12} recently studied four bright quasars with the Australia Telescope Compact Array (ATCA). They found that two of the four quasars are not well fitted with a model with a single component, and concluded that each quasar has multiple components with different RMs. This method is very powerful to interpret the source structure, especially in the case of multiple sources, when wideband data is available \citep{fa11}. Thus, we expect that the QU-fitting is also suited to the study of the IGMF in filaments of galaxies which is located between the radio source and the Galaxy or external galaxies along the LOS.

In this paper, we investigate how accurately the QU-fitting method can identify a weak IGMF such as that in filaments of galaxies with ongoing radio telescopes including the ASKAP, LOFAR, and GMRT. We consider an observation of a compact source such as quasars behind a diffuse source such as the Galaxy. RM due to the IGMF is naturally incorporated into the source model as a difference of RMs of the two radio sources. We employ the Fisher analysis to estimate the expected error in RM of the IGMF, accounting for the telescope's frequency coverage, channels, and sensitivities. Below, in section 2, we describe our model of polarization observations and procedure of the Fisher analysis. In section 3, we show the results of the Fisher analysis. Finally, we give a summary and discussion in section 4.

\section{Model and Calculation}
\label{section2}

\subsection{Basic Concepts}
\label{section2.1}

We start with defining key physical quantities in this paper. Hereafter, we consider the case that cosmological redshift is not significant.

The complex polarized intensity observed at a wavelength $\lambda$ can be written as,
\begin{equation}
P(\lambda^2)
= p(\lambda^2)I(\lambda^2)
= Q(\lambda^2)+iU(\lambda^2)
\end{equation}
where $p$ is the degree of polarization, and $I$, $Q$ and $U$ are the Stokes parameters. Following the same manner as that described by \cite{br05}, the polarized intensity can be written as
\begin{equation}\label{eqPI}
P(\lambda^2)
=\int_{-\infty}^{+\infty}F(\phi)e^{2i\phi\lambda^2} d\phi,
\end{equation}
where $F(\phi)$ is the Faraday dispersion function (FDF), which is a complex polarized intensity per unit Faraday depth, $\phi$. The Faraday depth (rotation measure, RM) to the source located at physical distance, $x$, is given by
\begin{equation}\label{Faraday depth}
\phi(x)
=810 \int_x^0 n_{\rm e} (x') B_\parallel(x')dx'~\frac{\rm rad}{\rm m^2},
\end{equation}
where $n_{\rm e}$ is the electron density in units of ${\rm cm^{-3}}$, $B_\parallel$ is the magnetic field strength along the LOS in $\mu$G, and $x'$ is the physical distance in kpc.

In the QU-fitting method,  we first suppose distribution of radio sources and magnetic fields,  then construct the corresponding model $F(\phi)$ with free parameters such as RM due to the IGMF. The free parameters are estimated by fitting the model  Q and U data with the observed  Q and U data. In this paper, instead of using the observed data, we investigate the expected errors on the fit by the Fisher analysis (Section 2.4).

Alternatively, $F(\phi)$ can be directly derived from polarization data $P(\lambda^2)$ by the inverse of Eq. (\ref{eqPI}),
\begin{equation}\label{eqF}
F(\phi)
=\int_{-\infty}^{+\infty}P(\lambda^2)e^{-2i\phi\lambda^2}d\lambda^2.
\end{equation}
This method is called Faraday tomography or Faraday RM Synthesis. Although we do not use Faraday tomography in this paper, Eq. (\ref{eqF}) will be useful for the QU-fitting method because it can be used to guess the reasonable $F(\phi)$ to be fitted and to check the consistency of the results of the fitting. Readers who have interests in the detailed explanations and applications of Faraday tomography should refer to recent works (e.g., \cite{br05,sch07,sch09,h09,hbe09,mao10,fr11,li11,andr11,aktr12} and references therein).

\subsection{Model FDF}
\label{section2.2}

\begin{figure}[tp]
\figurenum{1}
\epsscale{0.7}
\plotone{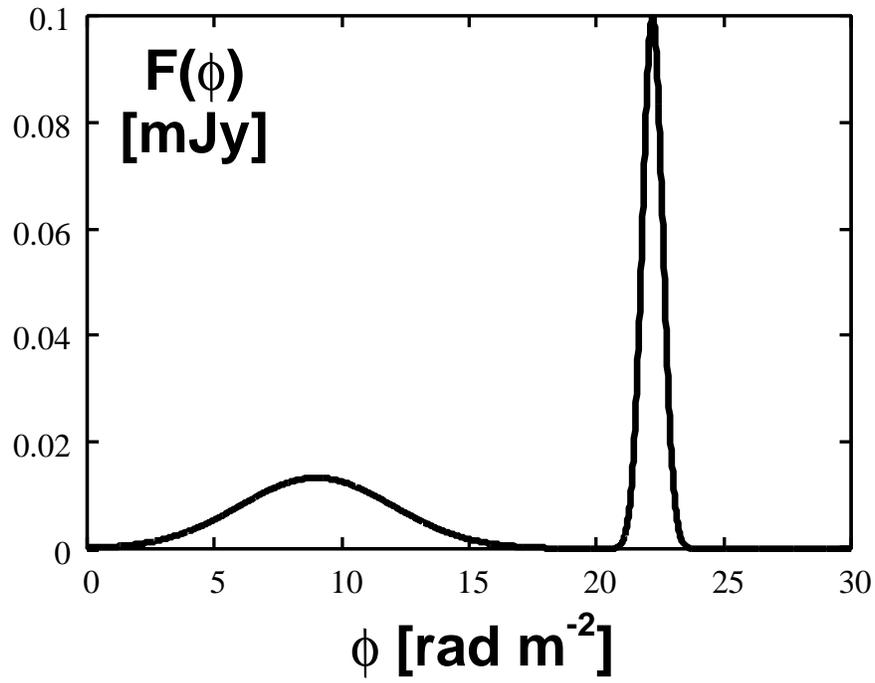}
\caption{Model Faraday dispersion function with, ($f_{\rm d}$, $\phi_{\rm d}$, $\delta \phi_{\rm d}$, $\theta_{\rm d}$, $f_{\rm c}$, $\phi_{\rm c}$, $\delta\phi_{\rm c}$, $\theta_{\rm c})$ $=$ ($0.1~{\rm mJy}$, $9.0~{\rm rad/m^2}$, $3.0~{\rm rad/m^2}$, $\pi/4~{\rm rad}$, $0.1~{\rm mJy}$, $22.2~{\rm rad/m^2}$, $0.4~{\rm rad/m^2}$, $0.0~{\rm rad}$). In this case, $RM_{\rm IGMF} = 3.0~{\rm rad/m^2}$.}
\label{fig1}
\end{figure}

We construct a model $F(\phi)$ as follows. We consider an observation of a compact source such as quasars through a diffuse source such as the Galaxy assuming the existence of the IGMF in the cosmic web. An example of the model FDF is shown in Fig. \ref{fig1}. We regard them as Faraday thick sources, which have finite thickness in $\phi$ space. Finite thickness is induced by magnetized plasma co-existing with the cosmic-rays emitting synchrotron radiation. The main results in this paper will be rather improved if Faraday thin (the thickness is small enough) sources are available, since they sharpen the gap (see below). 

If the IGMF exists between the diffuse and compact sources and the sign of the RM due to the IGMF is the same as those of the two sources, there is a gap between the two sources in $\phi$ space  (Fig. \ref{fig1}). This is because polarized intensity of the IGM is expected to be much smaller than those of the sources. Hence, if we obtain small errors in the model parameters enough to identify the gap, this means the detection of the IGMF and we can estimate its strength assuming the electron density of the IGM. Although the above situation that the signs of all RMs are the same does not always happen, we can choose such a source from many sources.

In fact, the gap between the two sources in $\phi$ space is also induced by intrinsic RMs of the sources, that is, RMs associated with the source but located out of the radio-emitting region, and RMs from a population of discrete intervening galaxies. These contributions shift the FDFs of the sources in $\phi$ space, which can cause a significant uncertainty in the estimation of the IGMF. However, effects of any intrinsic RMs could decrease by $(1+z)^{-2}$ if we observe high-redshift sources, and the LOS containing intervening galaxies could be excluded based on a tight correlation with optical absorption (see \cite{aktr12} for further discussions).

We assume a Gaussian shape of the sources in $\phi$ space, so that each source is characterized by the position, width, amplitude of polarized intensity, and the polarization angle. As will be discussed later, the Gaussian shape would not be realistic but this assumption makes analyses much simpler. Thus, the model FDF can be written as,
\begin{eqnarray}\label{FDFmodel}
F(\phi)
&=& \frac{f_{\rm d}}{\sqrt{2\pi} \delta\phi_{\rm d}} e^{2 i \theta_{\rm d}}
    \exp\left\{ -\frac{(\phi-\phi_{\rm d})^2}{2\delta\phi_{\rm d}^2} \right\}
\nonumber \\
&+& \frac{f_{\rm c}}{\sqrt{2\pi} \delta\phi_{\rm c}} e^{2 i \theta_{\rm c}}
    \exp\left\{ -\frac{(\phi-\phi_{\rm c})^2}{2\delta\phi_{\rm c}^2} \right\},
\end{eqnarray}
where $\phi_{\rm d}$ and $\phi_{\rm c}$ are the Faraday depths up to the centers of the sources in units of ${\rm rad/m^2}$, $\delta \phi_{\rm d}$ and $\delta \phi_{\rm c}$ are in ${\rm rad/m^2}$,  which determine the Faraday thickness of the source, $f_{\rm d}$ and $f_{\rm c}$ are the polarized intensities in mJy, and $\theta_{\rm d}$ and $\theta_{\rm c}$ are the initial polarization angles in radian. Here subscripts d and c represent the diffuse and compact sources, respectively. Therefore, our FDF model consists of total eight parameters.

To quantify the IGMF, we define the Faraday depth of the source as 3-$\sigma$ region from the center in $\phi$ space, and define RM due to the IGMF, $RM_{\rm IGMF}$, as
\begin{equation}\label{IGMF}
RM_{\rm IGMF}
= \left( \phi_{\rm c} - 3 \delta \phi_{\rm c} \right)
  - \left( \phi_{\rm d} + 3 \delta \phi_{\rm d} \right).
\end{equation}
\cite{ar10} estimated the rms value of RM through a single filament at the local universe to be $\sim 1~\rm{rad/m^2}$. Then, \cite{ar11} extended the study for filaments of galaxies up to the redshift of $z=5$, taking the redshift distribution of radio galaxies into account. They found that accumulation of RM through multiple filaments is a random walk process, and, up to $z=5$, the rms value reaches several $\rm{rad/m^2}$. Accounting for their results, below we consider RM of the IGMF in the range of ${\rm RM_{IGMF} = 1 - 3~rad/m^2}$.

\subsection{Specification of Radio Observatories}
\label{section2.3}

We investigate expected errors in $RM_{\rm IGMF}$ from future observations with ideal combinations of ongoing radio telescopes including ASKAP, GMRT and LOFAR. Table 1 shows the specifications of the telescopes. The adopted observatories cover wide frequency band from $\sim$100 MHz to $\sim$2 GHz, though there are a lack of data around $\sim$400-500 MHz. We do not use LOFAR LBA (30-80 MHz) in this study, since polarization is almost fully depolarized in this range and the inclusion of this lowest band data does not improve our results.

For a given number of channels, we divide $\lambda^2$ space evenly so as to divide the $\phi$ space evenly,  according to the conjugate relation between $\lambda^2$ and $\phi$. We notice that even sampling in $\lambda^2$ space in the observation is an important subject for future radio astronomy.

The sensitivity for the binned channel is calculated by reference to the specifications of these telescopes (Table 1) taking into account the fact that these sensitivities are calculated in frequency space, not in $\lambda^2$ space. For ASKAP and GMRT, we calculate the sensitivities from bandwidth of each channel ($B$) and system temperature ($T_{\rm sys}$) and total collecting area ($A_{\rm eff}$) of each telescope (Table 1) using the equation
\begin{equation}
S=\frac{k_{\rm B}T_{\rm sys}}{A_{\rm eff}\sqrt{Bt}},
\end{equation}
where $S$ is the sensitivity of each channel, $k_{\rm B}$ is the Boltzmann constant and $t$ is the observation time. For LOFAR, we calculate the sensitivities by converting sensitivities described in LOFAR website (Table 2) taking the bandwidths into account. We suppose one hour exposure in this paper.

\subsection{Fisher Analysis}
\label{section2.4}

Given observational data, the maximum likelihood analysis gives a set of model parameter values which maximizes the likelihood function. The expected errors in model parameters by this technique can be estimated by the Fisher analysis (e.g., \cite{coe09}). Fisher matrix is defined as
\begin{equation}
\mathcal F_{ij}
= \left. - \frac{\partial^2 \ln {\cal L}}{\partial p_i \partial p_j}
  \right|_{\vec{p} = \tilde{\vec{p}}}
= \left. \frac{1}{2}\frac{\partial^2\chi^2}{\partial p_i \partial p_j}
 \right|_{\vec{p} = \tilde{\vec{p}}},
\end{equation}
where $\cal L$ is the likelihood function, $p_i$ is the $i$-th model parameter, $\vec{p}=(p_1,p_2,\cdots)$ is a set of parameters and $\tilde{\vec{p}}=(\tilde{p}_1,\tilde{p}_2,\cdots)$ is the fiducial set of parameters around which confidence regions are put. In the last equation, we assumed a Gaussian likelihood and the Fisher matrix can be written with the Chi-squared value,
\begin{equation}
\chi^2(\vec{p})
=\sum_{l=1}^N \frac{\left[ Y_l(\vec{p}) - Y_l(\tilde{\vec{p}}) \right]^2}{\sigma_l^2},
\end{equation}
where $N$ is the number of data, $Y_l$ is the $l$-th data calculated from the model with the indicated set of parameters, and $\sigma_l$ is the observational error. Noting that the Fisher matrix is evaluated at $\vec{p} = \tilde{\vec{p}}$, it can be reduced to
\begin{equation}\label{eqFisher}
\mathcal F_{ij}
=\sum_{l=1}^N \frac{1}{\sigma_l^2}
\left[ \frac{\partial Y_l(\vec{p})}{\partial p_i}\frac{\partial Y_l(\vec{p})}{\partial p_j} \right]_{\vec{p} = \tilde{\vec{p}}}.
\end{equation}
In our case, observational data is the Stokes parameters, $Q$ and $U$, for each channel and telescope and by substituting them to $Y_l(p)$ in Eq. (\ref{eqFisher}), the Fisher matrix can be expressed as,
\begin{eqnarray}
\mathcal F_{ij}
=\sum_{l=1}^N
\left[
\frac{1}{{\sigma^2(\lambda^2_l)}}
\left\{
\frac{\partial Q_{\rm mod}(\lambda^2_l;\vec{p})}{\partial p_i}\frac{\partial Q_{\rm mod}(\lambda^2_l;\vec{p})}{\partial p_j} \right. \right. \nonumber \\
\left. \left. + \frac{\partial U_{\rm mod}(\lambda^2_l;\vec{p})}{\partial p_i}\frac{\partial U_{\rm mod}(\lambda^2_l;\vec{p})}{\partial p_j}
\right\}_{\vec{p} = \tilde{\vec{p}}}
\right],
\end{eqnarray}
where the $Q_{\rm mod}$ and $U_{\rm mod}$ are the Stokes parameters at the central wavelength $\lambda_l$ of the $l$th channel calculated from the assumed model.

From the Fisher matrix, the expected 1-$\sigma$ error of the $i$-th parameter which is marginalized over the other parameters is given by,
\begin{equation}\label{eqCov}
\sigma_{p_i}^2=\left( \mathcal F^{-1} \right)_{ii},
\end{equation}
and the covariance between the $i$-th and $j$-th parameters is given by,
\begin{equation}
\sigma_{ij}^2=\left(\mathcal F^{-1}\right)_{ij}.
\end{equation}
We will show the confidence region of the model parameters calculated with the above formulation. We note that the confidence regions obtained from the Fisher analysis sometimes spread into physically meaningless regions in the parameter space such as where $\delta \phi_{\rm d}$ is negative. This is because the shape of $\chi^2$ around the fiducial values are assumed to be quadratic in terms of the parameters. In these cases, we simply cut the confidence regions at the boundaries.

\section{Result}
\label{section3}

\subsection{Representative Case}
\label{section3.1}

We first show the result of the representative case with
$f_{\rm d}=0.1~{\rm mJy}$, $\phi_{\rm d}=9.0~{\rm rad/m^2}$, $\delta\phi_{\rm d}=3.0~{\rm rad/m^2}$, $\theta_{\rm d}=\pi/4~{\rm radian}$, $f_{\rm c}=0.1~{\rm mJy}$, $\phi_{\rm c} = 22.2~{\rm rad/m^2}$, $\delta\phi_{\rm c}=0.4~{\rm rad/m^2}$, and $\theta_{\rm c}=0.0~{\rm radian}$  (Fig. \ref{fig1}). The resultant $RM_{\rm IGMF}$ is $3.0~{\rm rad/m^2}$. Since the peak amplitude of the FDF is proportional to $f/\delta\phi$ (Eq. \ref{FDFmodel}), the compact source is brighter than the diffuse source for the same $f$ ($f_{\rm d}=f_{\rm c}$) but different $\delta\phi$ ($\delta\phi_{\rm d}>\delta\phi_{\rm c}$).

\begin{figure}[tp]
\figurenum{2}
\epsscale{0.7}
\plotone{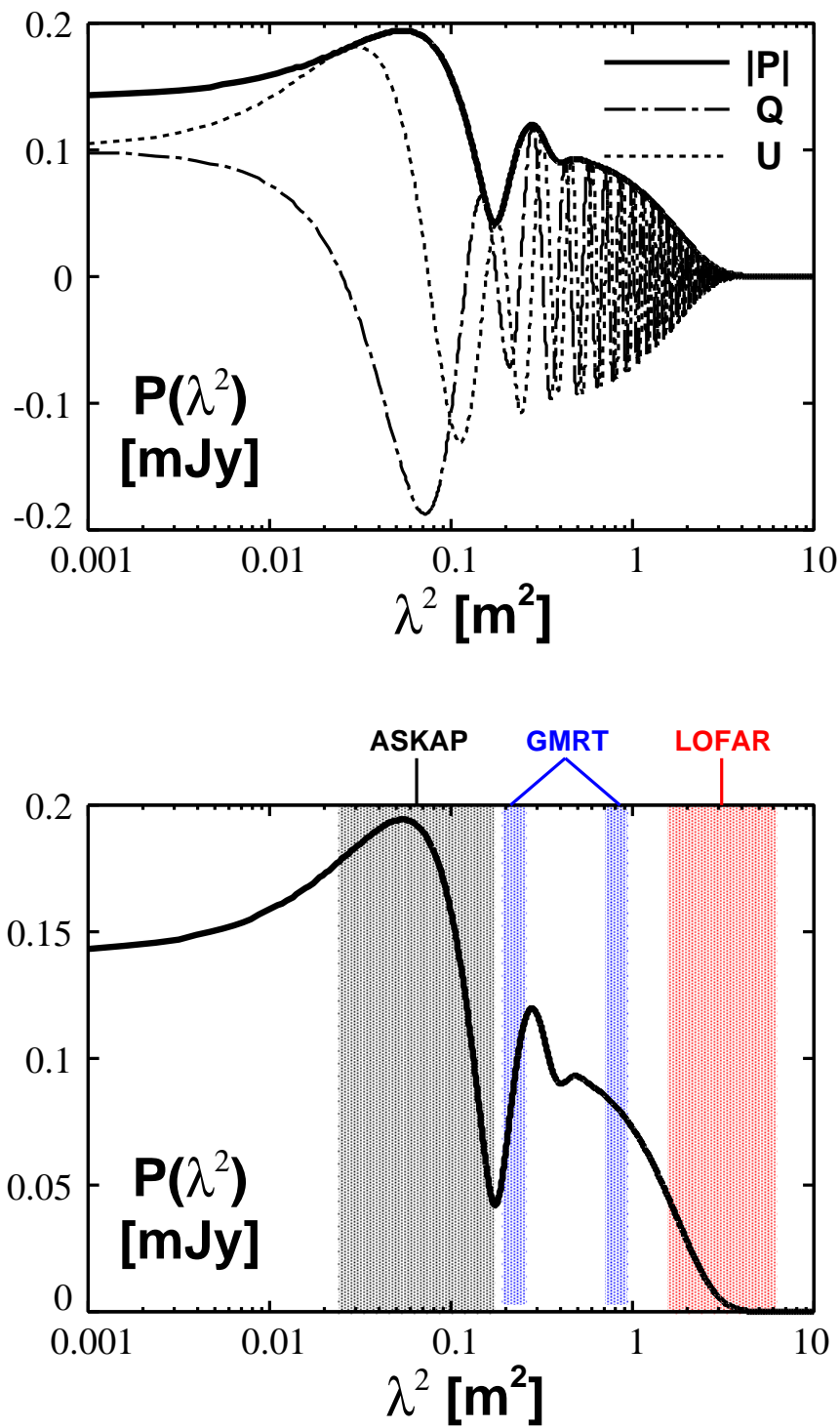}
\caption{Top panel: Polarized intensity $|P|$ (solid) and Stokes parameters $Q$ (dash-dotted) and $U$ (dotted), as functions of $\lambda^2$ which are calculated from the Faraday dispersion function shown in Fig. \ref{fig1}. Bottom panel: The band frequencies of ASKAP (black), GMRT (blue) and LOFAR (red), and polarization intensity $|P|$.}
\label{fig2}
\end{figure}

The top panel of Fig. \ref{fig2} shows the Stokes parameters $Q$ and $U$, and the polarized intensity $|P|$, calculated from the model FDF by Eq. (\ref{eqPI}). In the bottom panel of Fig. \ref{fig2}, the observation bands of ASKAP, GMRT and LOFAR are indicated as well as $|P|$. The value of $|P|$ globally decreases with increasing wavelength because of the Faraday depolarization, which is caused by  Faraday thickness of each source and is more significant for the diffuse source with a larger  thickness. There also exist wavelength-independent depolarization, which is caused by the interference between the two sources. As a result, the polarized intensity significantly drops to $O(10^{-2})~{\rm mJy}$ at the LOFAR band. Thus, the diffuse source can be mainly observed with ASKAP, while GMRT and LOFAR are sensitive to the compact source.

\begin{figure}[tp]
\figurenum{3}
\epsscale{0.9}
\plotone{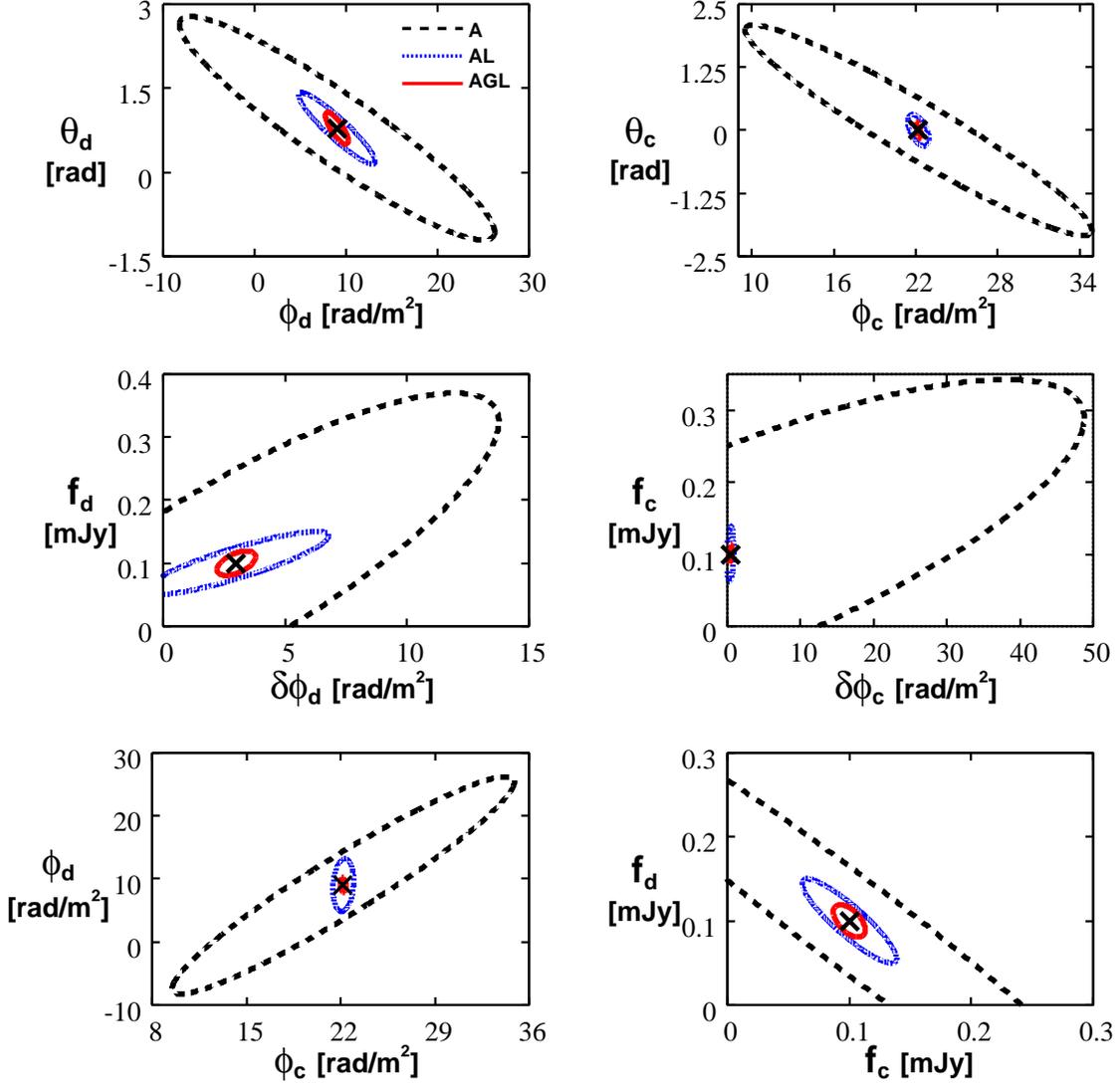}
\caption{Expected 1-$\sigma$ confidence ellipses for model parameters. Black dashed lines for ASKAP observation (A), blue dotted lines for ASKAP + LOFAR (AL) and red solid lines for ASKAP + GMRT + LOFAR (AGL). The assumed model parameters are the same as in Fig. \ref{fig1} and expressed by crosses.}
\label{fig3}
\end{figure}

Fig. \ref{fig3} shows the 1-$\sigma$ confidence ellipses obtained from the Fisher analysis for some selected pairs of model parameters. The black dashed, the blue dotted, and the red solid lines correspond to the results for ASKAP only (A) and the combinations of ASKAP + LOFAR (AL) and ASKAP + GMRT + LOFAR (AGL), respectively. The assumed parameters are indicated by crosses. We can see that pairs of parameters are more or less correlated each other, and the confidence region is improved when we consider combination observations. We only show combinations including ASKAP, because the confidence regions are much larger without ASKAP. This is related to the fact that the diffuse component  can be mainly observed with ASKAP. Also, the results for ASKAP + GMRT turned out to be almost the same as those for ASKAP + GMRT + LOFAR, and are not shown.

We find that $\delta\phi_{\rm c}$,  the parameter which determine the Faraday thickness of the compact source, is hardly determined by the observation with ASKAP alone. Because $\delta\phi_{\rm c}$ is one of the parameters which defines the IGMF (Eq. \ref{IGMF}), it implies that ASKAP itself cannot well constrain $RM_{\rm IGMF}$. This weak constraint by ASKAP is ascribed to the fact that short-wavelength observations cannot resolve small-scale structure in $\phi$ space. Actually, the parameter degeneracy and the constraint on $\delta \phi_{\rm c}$ are dramatically improved by the additions of longer-wavelength observations by GMRT and LOFAR.

\begin{figure}[tp]
\figurenum{4}
\epsscale{0.9}
\plotone{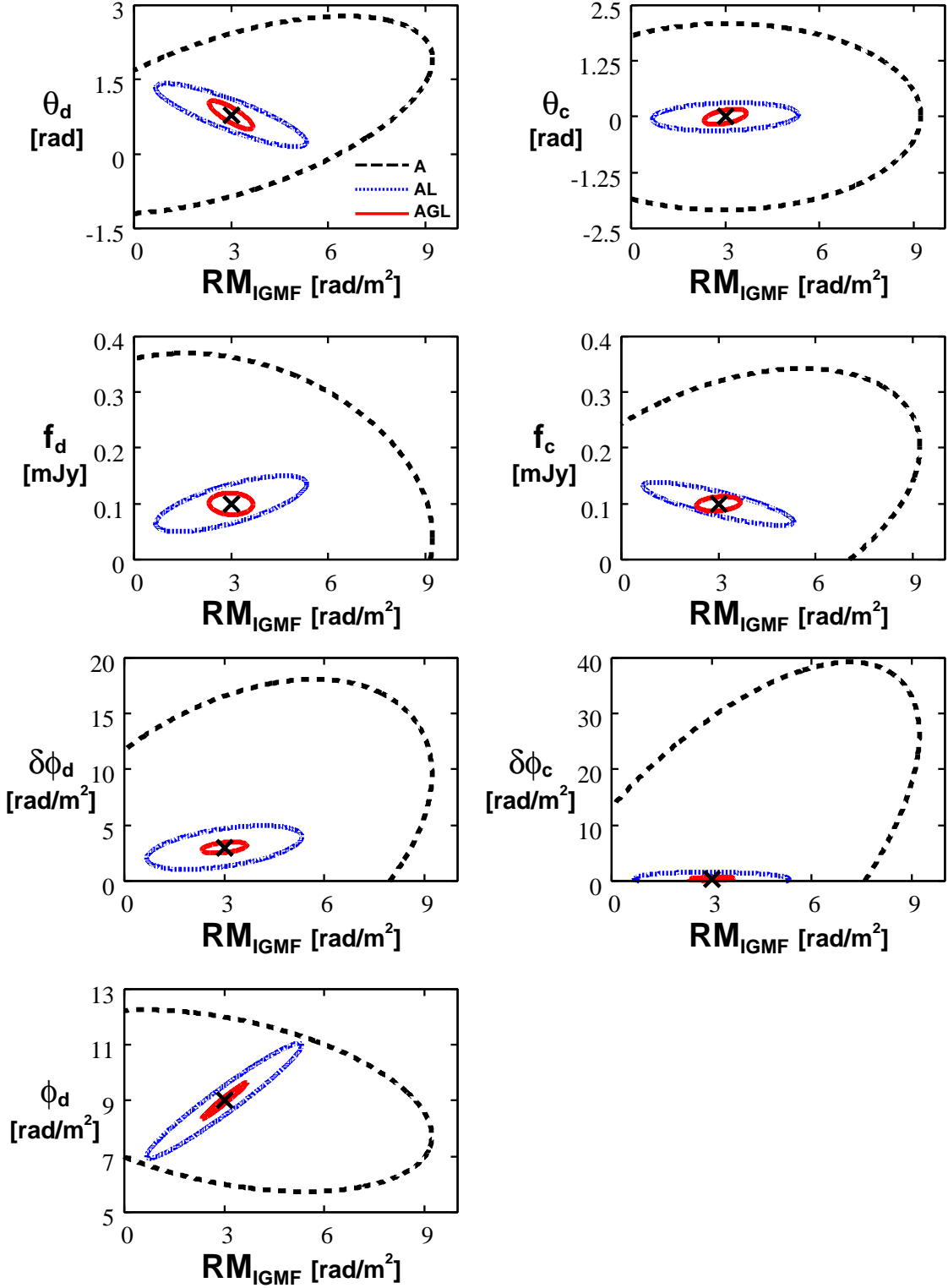}
\caption{Same as Figure \ref{fig3} but $\phi_{\rm c}$ is replaced with $RM_{\rm IGMF}$.}
\label{fig4}
\end{figure}

\begin{figure}[tp]
\figurenum{5}
\epsscale{1.0}
\plotone{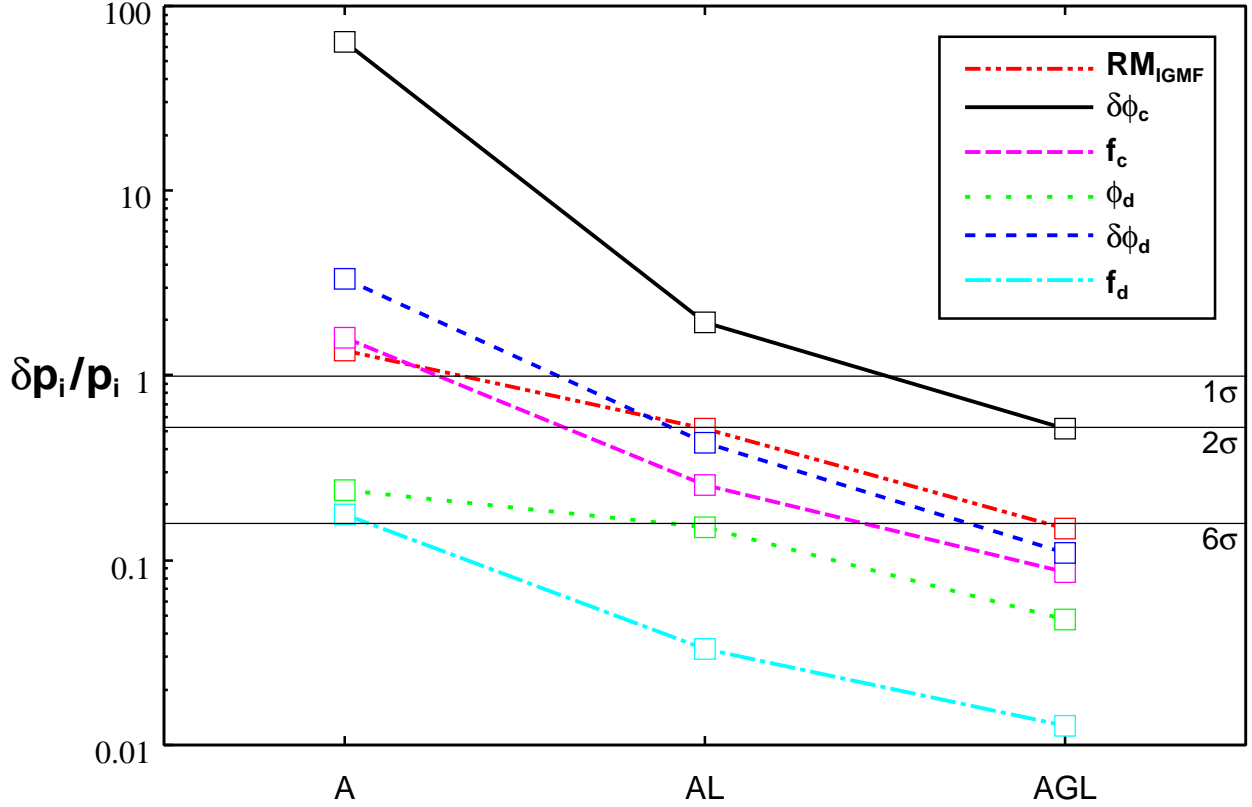}
\caption{Comparison of ratios of the marginalized 1-$\sigma$ errors to the parameter value in a logarithmic scale, $\log(\delta p_i/p_i)$, for various combinations of the telescopes. The three thin horizontal lines with labels 1, 2 and 6$\sigma$ are the significances to be able to constrain the parameters which correspond to $p_i/\delta p_i$. In the case of ${RM_{\rm IGMF}}$, these correspond to significances to be able to exclude zero IGMF ($RM_{\rm IGMF} = 0$). Labels A, AL and AGL represent ASKAP alone, ASKAP + LOFAR and ASKAP + GMRT + LOFAR, respectively.}
\label{fig5}
\end{figure}

To see how much RM due to the IGMF is constrained, we take $RM_{\rm IGMF}$ as an independent variable instead of $\phi_{\rm c}$. Fig. \ref{fig4} shows the 1-$\sigma$ confidence ellipses for model parameters with $RM_{\rm IGMF}$. Also the relative errors, which is the marginalized one-dimensional errors divided by the true values, are shown in Fig. \ref{fig5} for the three combinations of the telescopes. For ASKAP alone, the error is so large that zero IGMF ($RM_{\rm IGMF} = 0$) is not excluded at 1-$\sigma$ level. When we combine LOFAR, the situation drastically improves and zero IGMF can be excluded at about 2-$\sigma$ significance. By the full combination of ASKAP, LOFAR and GMRT, the significance increases up to about 6-$\sigma$. Thus, combination of these telescopes is very effective to probe the IGMF. Other parameters are also well determined by the combination of the telescopes, within $20\%$ for most of them. Only $\delta \phi_{\rm c}$ can not be determined well even by the combination. This is because of the lack of the sensitivity at long wavelengths where both sources become very dim due to the depolarization.

\subsection{Required Intensities to Detect the IGMF}
\label{section3.2}

We next consider general cases for various FDF models. We focus on some essential parameters to study the possibility of detecting the IGMF. Specifically, we examine the condition on the source intensities that the IGMF can be detected by our method.  We vary the three parameters, $f_{\rm d}, f_{\rm c}$ and $\phi_{\rm c}$, and fix the other five parameters, $\phi_{\rm d}=9.0~{\rm rad/m^2}$, $\delta\phi_{\rm d}=3.0~{\rm rad/m^2}$, $\theta_{\rm d}=0~{\rm radian}$, $\delta\phi_{\rm c}=0.4~{\rm rad/m^2}$, and $\theta_{\rm c}=\pi/4~{\rm radian}$ (same as the previous section). Varying $\phi_{\rm c}$ is equivalent to varying $RM_{\rm IGMF}$ for the fixed $\phi_{\rm d}$, $\delta\phi_{\rm d}$, and $\delta\phi_{\rm c}$, according to Eq. (\ref{IGMF}).

\begin{figure}[tp]
\figurenum{6}
\epsscale{0.8}
\plotone{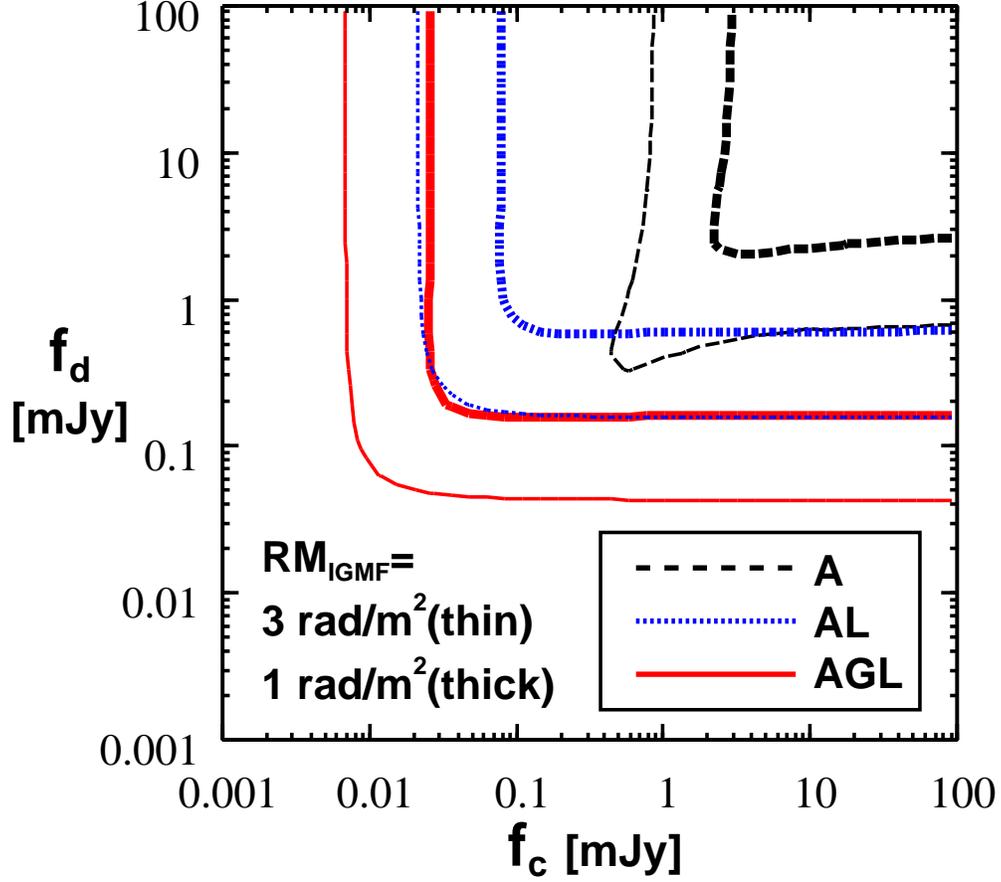}
\caption{Source intensities for the IGMF to be detected with 3-$\sigma$ significance. Black dashed lines are the intensities for ASKAP observation (A), blue dotted lines for ASKAP + LOFAR (AL) and red solid lines for ASKAP + GMRT + LOFAR (AGL). Two cases with $RM_{\rm IGMF} = 1.0~{\rm rad/m^2}$  (thick lines) and $3.0~{\rm rad/m^2}$  (thin lines) are shown. In the up-right regions of the lines, the IGMF is detected more than 3-$\sigma$ significance.}
\label{fig6}
\end{figure}

Fig. \ref{fig6} shows the regions on $f_{\rm c}$-$f_{\rm d}$ plane where non-zero IGMF is detected by 3-$\sigma$ significance for a  given $RM_{\rm IGMF}$ value, that is, 3-$\sigma$ error in $RM_{\rm IGMF}$ is smaller than the value of $RM_{\rm IGMF}$ itself. Two cases with $RM_{\rm IGMF} = 1.0$ and $3.0~{\rm rad/m^2}$ are plotted for each combination of the telescopes. Brighter sources are generally necessary for a smaller value of $RM_{\rm IGMF}$. In the case with $RM_{\rm IGMF} = 1.0~{\rm rad/m^2}$, we need much brighter (by a factor of ten) sources compared with the case with $RM_{\rm IGMF} = 3.0~{\rm rad/m^2}$ and the combination of the telescopes are very effective. On the other hand, in the case with $RM_{\rm IGMF} = 3.0~{\rm rad/m^2}$, even ASKAP alone can detect IGMF with relatively faint sources ($\sim 1~{\rm mJy}$). Considering the Galaxy as a diffuse source with an intensity,
\begin{equation}
I \sim 0.95
       \left( \frac{f}{\mathrm{1\;GHz}} \right)^{-1.5}
       \left( \frac{\Omega}{\mathrm{1\;arcmin^2}} \right)
       \rm{mJy},
\label{eqGalacticEmission}
\end{equation}
\citep{gol11} where $f$ is the frequency and $\Omega$ is the beam size, we need a compact source as bright as $\sim 20~{\rm \mu Jy}$ and $\sim 10~{\rm \mu Jy}$ for $RM_{\rm IGMF} = 1.0$ and $3.0~{\rm rad/m^2}$, respectively.

\section{Discussion and Summary}
\label{section4}

In this paper, we studied the capability to probe the intergalactic magnetic field (IGMF) by radio polarimetry using ongoing radio telescopes such as ASKAP, GMRT, and LOFAR. We considered observations of a compact radio source such as quasars through a diffuse source such as the Galaxy and constructed a model of Faraday dispersion function along a line of sight, where the IGMF appears as a gap between the two sources. We performed the Fisher analysis accounting for frequency  coverages, number of channels, and the sensitivities of these telescopes, and evaluated errors in the RM of the IGMF expected by future observation with the three telescopes. We examined the condition on the source intensities which enable us to detect the IGMF. It was shown that the IGMF with the rotation measure as small as $3.0~{\rm rad/m^2}$ can be detected by observing a compact source with intensities of $20~{\rm \mu Jy}$ with the combination of the three telescopes.

The above result is obtained in the best case and in reality there could be a number of the factors which introduce additional uncertainties on the fit, such as the polarization purity and errors of the polarization calibration as well as time variabilities of the ionospheric contamination and the radio frequency interference. Furthermore, in our analyses, it is critical to assume a specific shape of the Faraday dispersion function of the sources, as we adopted the Gaussian shape. Although the Gaussian, top-hat and delta-function shapes have conventionally been used in the literature \citep{bu66,sok98,osu12,aktr12}, the source shape, especially galactic one, would actually be more complicated. If this is the case, the boundary of the source may be more ambiguous, which makes it rather difficult to probe IGMFs because they are identified as a gap between two sources. Consequently, constraints on model parameters including the RM of the IGMF would become weaker in a realistic situation. Nevertheless, the results obtained here would give a reasonable estimate of the parameter errors if the real Faraday dispersion functions are relatively simple with small skewness. To be more realistic, we will need galaxy models which simulate the distribution of global and turbulent magnetic fields and high-energy electrons which emit synchrotron radiation. Study of realistic Faraday dispersion function of galaxies and the possibility of detecting IGMFs with it are currently ongoing and will be presented elsewhere.

Combination observations with ASKAP, GMRT, and LOFAR can cover wide frequency bands from $\sim$100 MHz to $\sim$2 GHz. But there are a lack of data around $\sim$400-500 MHz. We expect that the QU-fitting method would provide much better results for future SKA observations with seamless, ultra-wide frequency coverage. Note that our analyses are easily applied to other telescopes. Further, we can also consider a compact source behind an external galaxy as well as the Galaxy. In this case, the Faraday dispersion function has three components including the Galaxy and the IGMFs would appear as gaps between the sources depending on their configuration. Thus, targeting a compact source behind a bright external galaxy would enhance the possibility of detecting IGMFs.

To get an initial guess of model Faraday dispersion function in actual observations, Faraday tomography will be useful. Although limited frequency bands give us only its imperfect reconstruction, RMCLEAN and/or other method which improve the reconstruction will allow us to develop a more reasonable dispersion function \citep{h09,bel12}. Development in this direction is also of crucially importance to find the IGMF in filaments of galaxies.

\acknowledgments

This work is supported in part by the Grant-in-Aid from the Ministry of Education, Culture, Sports, Science and Technology (MEXT) of Japan, No. 23740179, No. 24111710 and No. 24340048 (KT). T.A. acknowledges the supports of the Japan Society for the Promotion of Science (JSPS).


\begin{deluxetable}{lcccc}
\tablenum{1}
\tabletypesize{\footnotesize}
\tablewidth{0pt}
\tablecaption{Specifications of radio observatories.\label{table1}}
\tablehead{
\colhead{Observatory} & 
\colhead{Frequency} &
\colhead{System temperature} &
\colhead{Effective area} &
\colhead{Number of channels} \\
& (GHz) & (K) & (${\rm m^2}$)
}
\startdata
LOFAR HBA\tablenotemark{a} & 0.120--0.240 & - & - & 156,000 \\
GMRT 327\tablenotemark{b} & 0.305--0.345 & 108 & 30000 & 256 \\
GMRT 610\tablenotemark{b} & 0.580--0.640 & 102 & 30000 & 256\\
ASKAP\tablenotemark{c} & 0.700--1.800 & 50 & 4072 & 60,000
\enddata
\tablenotetext{a}{http://www.astron.nl/radio-observatory/astronomers}
\tablenotetext{b}{\cite{ana95}}
\tablenotetext{c}{http://www.atnf.csiro.au/projects/mira/specs.html}
\end{deluxetable}

\begin{deluxetable}{cc}
\tablenum{2}
\tabletypesize{\footnotesize}
\tablewidth{0pt}
\tablecaption{Sensitivity of the LOFAR array with bandwidth of 3.57 MHz\tablenotemark{*}.\label{table2}}
\tablehead{
\colhead{Frequency} & 
\colhead{Sensitivity} \\
 (MHz) & (mJy)
}
\startdata
120 & 0.20 \\
150 & 0.16 \\
180 & 0.18 \\
200 & 0.20 \\
210 & 0.21 \\
240 & 0.23
\enddata
\tablenotetext{*}{http://www.astron.nl/radio-observatory/astronomers/lofar-imaging-capabilities-sensitivity/sensitivity-lofar-array/sensiti}
\end{deluxetable}

\clearpage


\placefigure{f1}

\placefigure{f2}

\placefigure{f3}

\placefigure{f4}

\placefigure{f5}

\placefigure{f6}

\end{document}